\documentstyle[11pt]{article}
\input{psfig}

\newtheorem{theorem}{\sc Theorem}
\newtheorem{lemma}{\sc Lemma}
\newtheorem{coro}{\sc Corollary}
\newtheorem{nota}{\sc Notation}
\newtheorem{defin}{\sc Definition}
\newtheorem{rem}{\sc Remark}
\newtheorem{cla}{\sc Claim}
\newtheorem{ex}{\sc Example}

\newenvironment{proof}{\par \sc Proof.\rm}{\hspace*{\fill}$\Box$\vspace{1ex}}

\newenvironment{claim}{\begin{cla}}{\end{cla}}
\newenvironment{corollary}{\begin{coro}}{\end{coro}}

\newenvironment{remark}{\begin{rem}}{\end{rem}}

\title{Average-Case Complexity of Shellsort}
\author{Tao Jiang\thanks{Supported in part by the NSERC
Research Grant OGP0046613 and a CITO grant.
Address: Department of Computing and Software,
McMaster University, Hamilton, Ont L8S 4K1, Canada.
Email: jiang@cas.mcmaster.ca}\\
McMaster University
\and
Ming Li\thanks{
Supported in part by
the NSERC Research Grant OGP0046506, a CITO grant, and the
Steacie Fellowship. Address:
Department of Computer Science, University of Waterloo,
Waterloo, Ont. N2L 3G1, Canada. E-mail: mli@math.uwaterloo.ca}\\
University of Waterloo
\and
Paul Vit\'{a}nyi\thanks{Partially supported by the European Union
through NeuroCOLT II ESPRIT Working Group.
Address: CWI, Kruislaan 413, 1098 SJ Amsterdam, The Netherlands.
Email: paulv@cwi.nl}\\
CWI and University of Amsterdam}
\date{}
 
\begin{document}
\maketitle
 
\begin{abstract}
We prove a general lower bound on the average-case complexity of
Shellsort: the average number of
data-movements (and comparisons) made by a $p$-pass Shellsort
for any incremental sequence is $\Omega (pn^{1 + \frac{1}{p}})$ for 
all $p \leq \log n$.
Using similar arguments, we analyze the average-case complexity of
several other sorting algorithms.
\end{abstract}

\section{Introduction}
The question of a nontrivial general lower bound 
(or upper bound) on the average complexity of Shellsort
(due to D.L. Shell \cite{Sh59})
has been open for about four decades \cite{Kn73,Se97}.
We present such a lower bound for $p$-pass Shellsort
for every $p$. 

Shellsort sorts a list of $n$ elements in
$p$ passes using a sequence of increments
$h_1 , \ldots , h_p$. In the $k$th pass the
main list is divided in $h_k$ separate sublists
of length $n/h_k$,
where the $i$th sublist consists of the elements at positions
$i \bmod h_k$  of the main list ($i=1, \ldots , h_k$).
Every sublist is sorted using a straightforward insertion sort.
The efficiency of the method is governed
by the number of passes $p$ and
the selected increment sequence $h_1 , \ldots , h_p$
with $h_p =1$ to ensure sortedness of the final list.
The original $\log n$-pass increment sequence 
\footnote{``$\log$'' denotes the binary logarithm and ``$\ln$''
denotes the natural logarithm.
}
$\lfloor n/2 \rfloor , \lfloor n/4 \rfloor, \ldots , 1$ 
of Shell \cite{Sh59} uses worst case $\Theta (n^2)$ time,
but Papernov and Stasevitch \cite{PS65} showed that 
another related sequence uses $O(n^{3/2})$ and 
Pratt \cite{Pr72} extended this to a class of all nearly
geometric increment sequences and proved this bound 
was tight.
The currently best asymptotic method was found
by  Pratt \cite{Pr72}. It uses
all $\log^2 n$ increments of the form 
$2^i 3^j < \lfloor n/2 \rfloor$
to obtain time $O(n\log^2 n)$ in the worst
case. Moreover, since every pass
takes at least $n$ steps, the average complexity 
using Pratt's increment sequence is $\Theta (n \log^2 n)$.
Incerpi and Sedgewick \cite{IS85} constructed a family of
increment sequences for 
which Shellsort runs in $O(n^{1+ \epsilon / \sqrt{\log n}})$ time
using $(8/\epsilon^2) \log n$ passes,
for every $\epsilon > 0$.  B. Chazelle (attribution in  \cite{Se96})
obtained the same result by generalizing V. Pratt's method:
instead of using 2 and 3 to construct the increment sequence
use $a$ and $(a+1)$ for fixed $a$ which yields a 
worst-case running time of $n \log^2 n (a^2/\ln^2 a)$
which is $O(n^{1+ \epsilon / \sqrt{ \log n}})$ for $\ln^2 a = O(\log n)$.
Plaxton, Poonen and Suel \cite{PPS92} proved
an $\Omega (n^{1 + \epsilon / \sqrt{p}} )$
lower bound for $p$ passes of Shellsort using 
any increment sequence, for some $\epsilon > 0$; 
taking $p = \Omega (\log n)$ shows that the Incerpi-Sedgewick / Chazelle
bounds are optimal for small $p$ and taking $p$ slightly larger
shows a $\Theta ( n \log^2 n / (\log \log n)^2)$ lower bound
on the worst case complexity of Shellsort.
Since every pass takes at least $n$ steps this shows
an $\Omega (n \log^2 n/ (\log \log n)^2)$ lower bound
on the worst-case of every Shellsort increment sequence.
For the {\em average-case} running time Knuth \cite{Kn73} shows
$\Theta ( n^{5/3})$ for the best choice of increments in $p=2$ passes
and Yao \cite{Yao80} derives an expression for
the average case for $p=3$ that doesn't result in a comparable asymptotic
analytic bound. Apart from this no nontrivial results
are known for the average case; see \cite{Kn73,Se96,Se97}.

\bigskip
{\noindent \bf Results:} We show a general $\Omega (p n^{1+1/p})$ lower
bound on the average-case running time of $p$-pass Shellsort
under uniform distribution of input permutations for $p \leq \log n$. 
For $p > \log n$ the lower bound is trivially $\Omega (pn)$. This is the first
advance on the problem 
of determining general nontrivial bounds on the {\em average-case} running time
of Shellsort~\cite{Pr72,Kn73,Yao80,IS85,PPS92,Se96,Se97}. 
Using the same simple method,
we also obtain results on the average number of stacks or queues
(sequential or parallel) required for sorting under the uniform
distribution on input permutations. These problems have been 
studied before by Knuth \cite{Kn73} and Tarjan \cite{T72} for the
worst case.

\bigskip
{\noindent \bf Kolmogorov complexity and the Incompressibility Method:}
The technical tool to obtain our results
is the incompressibility method.
This method is especially suited for the
average case analysis of algorithms and machine models,
whereas average-case analysis is usually more difficult than worst-case
analysis using more traditional methods.
 A survey of the use of the incompressibility
method is \cite{LiVi93} Chapter 6, and recent work is \cite{BJLV99}.
The most spectacular
successes of the method
occur  in the computational complexity analysis of algorithms.

Informally, the Kolmogorov
complexity $C(x)$ of a binary string~$x$ is the length of the shortest
binary program (for a fixed reference universal machine) that prints $x$
as its only output and then halts \cite{Ko65}. A string $x$ is {\em
incompressible\/} if $C(x)$ is at least $|x|$, the approximate length of
a program that simply includes all of $x$ literally.  
%
     Similarly, the {\em conditional\/} Kolmogorov complexity of $x$
with respect to $y$, denoted by $C(x|y)$, is the length of the shortest
program that, {\emph with extra information $y$}, prints $x$.  And a
string $x$ is incompressible {\emph relative to
$y$} if $C(x|y)$ is large in the appropriate sense.
For details see \cite{LiVi93}. 
Here we use that, both absolutely and
relative to any fixed string $y$, there are incompressible strings of
every length, and that {\em most\/} strings are nearly incompressible,
by {\em any\/} standard.
\footnote{
By a simple counting argument one can show
that whereas some strings can be enormously compressed,
like strings of the form $11 \ldots 1$,
the majority of strings can hardly be compressed at all.
For every $n$ there are $2^n$ binary
strings of length $n$, but only
$\sum_{i=0}^{n-1} 2^i = 2^n -1$ possible shorter descriptions.
Therefore, there is at least one binary string
$x$ of length $n$ such that $C(x) \geq n$.
Similarly,
for every length $n$ and any binary string $y$,
there is a binary string $x$ of length $n$ such that
$C(x| y)   \geq   n$.}
Another easy one is that significantly long
subwords of an incompressible string are
themselves nearly incompressible
by {\em any\/} standard,
even relative to the rest of the string.
\footnote{Strings that are incompressible
are patternless,
since a pattern could be used to reduce
the description length. Intuitively, we
think of such patternless sequences as being random, and we
use ``random sequence'' synonymously with ``incompressible sequence.''
It is possible to give a rigorous formalization of the intuitive notion
of a random sequence as a sequence that passes all
effective tests for randomness, see for example \cite{LiVi93}.}
In the sequel we use the following easy facts (sometimes only implicitly).
\begin{lemma}
\label{C2}
Let $c$ be a positive integer.
For every fixed $y$, every
finite set $A$ contains at least $(1 - 2^{-c} )|A| + 1$
elements $x$ with $C(x| A,y)   \geq   \lfloor \log |A| \rfloor  - c$.
\end{lemma}

\begin{lemma}
\label{C3}
If $A$ is a set, then for every $y$ 
every element $x \in A$ has complexity
$C(x| A,y) \leq \log |A| + O(1)$.
\end{lemma}
The first lemma is proved by simple counting.
The second lemma holds since $x$ can be
described by first describing $A$ in
$O(1)$ bits and then giving the index of $x$
in the enumeration order of $A$.

\section{Shellsort}
A Shellsort computation consists of a sequence 
comparison and inversion (swapping) operations. 
In this analysis of the average-case lower bound we count just
the total number of data movements (here inversions) executed.
The same bound holds for number of comparisons automatically.

\begin{theorem}\label{theo.shelllb}
A lower bound on the average number of inversions in a 
p-pass Shellsort with $p \leq \log n$ is 
$
\Omega \left( pn^{1+\frac{1}{p}}
\right)$.
\end{theorem}

\begin{proof}
Let the list to be sorted consist of a permutation
$\pi$ of the elements $1, \ldots , n$.
Consider a  $(h_1 , \ldots , h_p)$ Shellsort where
$h_k$ is the increment in the $k$th pass and $h_p=1$.
We assume that $p \leq \log n$.
For any $1 \leq i \leq n$ and $1 \leq k \leq p$, let $m_{i,k}$ be 
the number of elements in the {\em $h_k$-chain} containing element $i$
that are to the left of $i$ at the beginning of pass $k$  
and are larger than $i$.
Observe that $\sum_{i=1}^{n} m_{i,k}$ is the number of inversions
in the initial permutation of pass $k$, and that the insertion sort
in pass $k$ requires precisely $\sum_{i=1}^{n} (m_{i,k} +1)$
comparisons.
Let $M$ denote the total number of inversions:
\begin{equation}\label{eq.M}
 M := \sum_{i,k=1}^{n,p} m_{i,k}. 
\end{equation}

\begin{claim}\label{lem.descr}
Given all the  $m_{i,k}$'s in an appropriate fixed order,
we can reconstruct the original permutation $\pi$.
\end{claim}
\begin{proof}
The $m_{i,p}$'s trivially specify the initial permutation of pass $p$.
In general, given the $m_{i,k}$'s and the final permutation of pass $k$,
we can easily reconstruct the initial permutation of pass $k$.
\end{proof}

Let $M$ as in (\ref{eq.M}) be a fixed number. Let permutation $\pi$ be a random
permutation having Kolmogorov complexity
\begin{equation}\label{eq.compl}
C(\pi | n,p,P) \geq \log n! - \log n.
\end{equation}
where $P$ is the encoding program in the following discussion.
The description in Claim~\ref{lem.descr} is effective
and therefore its minimum length must exceed the complexity of $\pi$:
\begin{equation}\label{eq.excC}
C(m_{1,1} , \ldots , m_{n,p} | n,p,P)
\geq C( \pi |n ,p,P) .
\end{equation}
Any $M$ as defined by (\ref{eq.M}) such that every
division of $M$ in $m_{i,k}$'s contradicts (\ref{eq.excC})
would be a lower bound on the number of inversions performed.
There are 
\begin{equation}\label{eq.DM}
 D(M) := \sum_{i=1}^{np-1} {M \choose {np-i}} 
= {{M+np-1} \choose {np-1}}.
\end{equation}
possible divisions of $M$
into $np$ nonnegative integral summands $m_{i,k}$'s. Every
division can be indicated by its index $j$ in an enumeration
of these divisions. Therefore, a self-delimiting description of $M$
followed by a description of $j$ effectively describes the $m_{i,k}$'s.
The length of this description must by definition exceed the
length of the minimal effective description (the Kolmogorov complexity).
That is,
\[ \log D(M) + \log M + 2 \log \log M  \geq
C(m_{1,1} , \ldots , m_{n,p} | n, p, P) + O(1).\]
We know that $M\leq pn^2$ since every $m_{i,k} \leq n$.
We also don't need to consider $p = \Omega (n)$.
Together with (\ref{eq.compl}) and (\ref{eq.excC}), we have
\begin{equation}\label{eq.Mnp}
\log D(M) \geq \log n! - 4 \log n+ O(1) .
\end{equation}

{\bf Case 1:} Let $M \leq  np-1$. Then 
\[\log D(M) \leq \log {M \choose M/2} = 
M - \frac{1}{2} \log M .
\]
Using (\ref{eq.Mnp}) we find
$M = \Omega (n \log n )$ and $p = \Omega ( \log n )$.  

{\bf Case 2:} Let $M \geq np$. Then
by (\ref{eq.DM}) $D(M)$ is bounded above by \footnote{
Use the following formula (\cite{LiVi93}, p. 10), 
\[ \log {a \choose b} = b \log \frac{a}{b} + (a-b) \log \frac{a}{a-b}
+ \frac{1}{2} \log \frac{a}{b(a-b)} + O(1) .\]
}
\begin{eqnarray*}
 \log {{M+np-1} \choose {np-1}}  =  && (np-1) \log \frac{M+np-1}{np-1}
 +M\log \frac{M+np-1}{M} \\
&& + \frac{1}{2} \log \frac{M+np-1}{(np-1)M} + O(1). 
\end{eqnarray*}
The second term in the right-hand side equals
\[ \log \left( 1+ \frac{np-1}{M} \right)^{M}
\rightarrow \log e^{np-1} \]
for $n \rightarrow \infty$. The third term in the righthand side
goes to 0 for $n \rightarrow \infty$. Therefore,
the total right-hand side goes to
\[ (np-1) \left( \log \frac{M+np-1}{np-1}+ \log e \right) \]
for $n \rightarrow \infty$. Together with (\ref{eq.Mnp}) this yields
\[ M = \Omega ( p n^{1+ \frac{1}{p}} ) . \]
Therefore, the running time of the algorithm is as
stated in the theorem for every permutation $\pi$ 
satisfying (\ref{eq.compl}). 
\footnote{Let us refine the argument by taking
into account the different increments
$h_1, \ldots ,h_p$ of the different passes
maximizing the contribution of every pass separately. Fix 
the number of inversions in pass $k$ as 
$M_k \leq n^2/h_k$ ($k := 1, \ldots, p$).
Replace $M$ in (\ref{eq.M}), (\ref{eq.compl}), and  (\ref{eq.excC})
by the vector $(M_1, \ldots , M_p)$. 
With $p \leq \log n$ 
encoding the $M_i$'s self-delimiting takes at 
most $p ( \log n^2 + 2 \log \log n^2) = O( \log^2 n )$ bits.  
If all 
$M_i > n$ ($1 \leq i \leq p$), then we find similar to before 
\[
\log {{M_1 + n-1} \choose {n-1}} \ldots 
{{M_p + n-1} \choose {n-1}} +  O( \log^2 n ) \geq  \log n! - 4 \log n + O(1).
\]
Altogether this leads to 
$\log ((M_1 + n-1)/(n-1)) \ldots ((M_p + n-1)/(n-1)) 
 =  \log n  - O ((\log^2 n)/n)$
which by the inequality of arithmetic and geometric means
($(\sum M_i ) /p \geq (\prod M_i )^{1/p}$)
yields $(M_1 + \ldots M_p +p(n-1))/p \geq n^{1+\frac{1}{p}}$.
Just like before we now obtain 
$M = M_1 + \ldots + M_p = \Omega (pn^{1+\frac{1}{p}})$. 
So we need some more subtle argument to improve the lower bound.
}
By lemma~\ref{C2} at least
a $(1-1/n)$-fraction of all permutations $\pi$ require
that high complexity. Therefore, the following is a
lower bound on the expected number
of comparisons of the sorting procedure:
\[ (1- \frac{1}{n}) \Omega ( p n^{1+ \frac{1}{p}} ) , \]
where we can ignore the contribution 
of the remaining $(1/n)$-fraction of all
permutations. This gives us the theorem.
\end{proof}

\begin{corollary}
\rm
We can do the same analysis for the number of comparisons. Denote
the analogues of $m_{i,k}$ by $m'_{i,k}$ and the analogue
of $M$ by $M'$. Note that $m'_{i,k} = m_{i,k}+1$.
Counting the number of comparisons we observe that every element
is compared at least once in every pass. Therefore, all the $m'_{i,k}$'s
are positive integers and the number $D(M) = {M \choose {np-1}}$.
A similar calculation yields that $M' = \Omega (pn^{1+1/p})$
again.
\end{corollary}

Compare our lower bound on the average-case with the Plaxton-Poonen-Suel 
$\Omega (n^{1 + \epsilon / \sqrt{p} } )$ worst case lower bound.
Some special cases of the lower bound on the average-case complexity are:

\begin{enumerate} 
\item
When $p=1$, this gives asymptotically tight bound for the 
average number of inversions for Insertion Sort.

\item When $p=2$, Shellsort requires $\Omega (n^{3/2} )$ inversions
(the tight bound is known to be $\Theta (n^{5/3})$ \cite{Kn73});

\item When $p=3$, Shellsort requires $\Omega (n^{4/3} )$ inversions
(\cite{Yao80} gives an analysis but not a comparable asymptotic
formula);


\item When $p= \log n / \log \log n$, Shellsort requires
$\Omega (n \log^2 n / \log \log n )$ inversions; 

\item When $p=\log n$, Shellsort requires $\Omega (n \log n)$
inversions. When we consider comparisons, 
this is of course the lower bound of average number of 
comparisons for every sorting algorithm.

\item When $p= \log^2 n$, 
Shellsort requires $\Omega (n \log n)$ inversions but it also
requires $\Omega (n \log^2 n)$ comparisons.
(The running time is known to be $\Theta (n \log^2 n)$ in this case
 \cite{Pr72}).

\item In general, when $p = p(n) > \log n$, 
Shellsort requires $\Omega (n \cdot p(n))$ comparisons
because every pass trivially makes $n$ comparisons.

\end{enumerate} 
In \cite{Se97} it is mentioned that the existence of an increment
sequence yielding an average $O(n \log n)$ Shellsort has been
open for 30 years. The above lower bound on the average shows
that the number $p$ of passes of such an increment sequence (if it exists)
is precisely $p=\Theta (\log n)$; all
the other possibilities are ruled out.

\begin{remark}
\rm
It is a simple consequence of the Shellsort analysis to obtain
average-case lower bounds on some other sorting methods. Here we
use Bubble Sort as an example. In the next section, we
analyze stack-sort and queue-sort.
A description and average-case analysis of Bubble Sort can be found in
\cite{Kn73}.
It is well-known that Bubble Sort uses $\Theta (n^2)$
comparisons/exchanges on the average. We present a very simple proof
of this fact.
The number of exchanges is obviously at most $n^2$, so we only 
have to consider the lower bound.
In Bubble Sort we make at most $n-1$ passes from
left to right over the permutation to be sorted and
move the largest element we have currently found
right by exchanges. For a permutation
$\pi$ of the elements $1, \ldots , n$, we can describe the total
number of exchanges by $M := \sum_{i=1}^{n-1} m_i$ where
$m_i$ is the initial distance of element $n-i$ to its proper place $n-i$.
Note that in every pass more than one element may ``bubble'' right but
that means simply that in the future passes
of the sorting process an equal number of exchanges will be saved
for the element to reach its final position. 
That is, every element executes a number of
exchanges going right that equals precisely the initial
distance between its start position to its final position.
An almost identical analysis as that of Theorem~\ref{theo.shelllb}
shows that $\log M/n \geq \log n + O(1)$
for every $M$. As before this holds for an overwhelming fraction
of all permutations, and hence
gives us an $\Omega (n^2)$ lower bound on the
expected number of comparisons/exchanges.
\end{remark}

\section{Sorting with Queues and Stacks}

\begin{figure}[thbp]
\hfill\ \psfig{figure=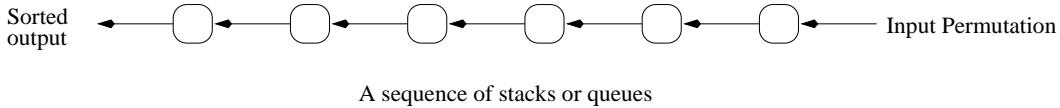,width=5.5in} \hfill\
\caption{Six stacks/queues arranged in sequential order}
\label{stacks}
\end{figure}

\begin{figure}[thbp]
\hfill\ \psfig{figure=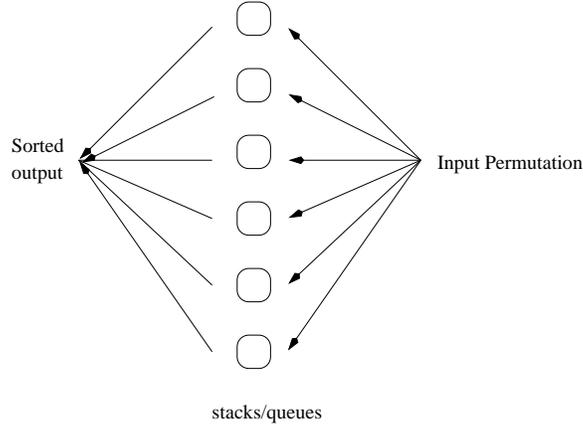,width=3in} \hfill\
\caption{Six stacks/queues arranged in parallel order}
\label{parallel}
\end{figure}

Knuth \cite{Kn73} and Tarjan \cite{T72} have studied the 
problem of sorting using a network of queues or stacks.
In particular, the main variant of the problem is: assuming the
stacks or queues are arranged sequentially as shown in
Figure~\ref{stacks} or in parallel as shown in
Figure~\ref{parallel}, then how many stacks or queues are needed to sort $n$
numbers. Here, the input sequence is scanned from left to right and
the elements follow the arrows to go to the next stack or queue or output.

Tarjan \cite{T72} actually studied arbitrary acyclic networks of stacks
and queues. Our technique will in general apply there. However we will
concentrate on dealing with the above two main variants, and concentrate on
the average-case analysis. 

\subsection{Sorting with Sequential Stacks}\label{sect.seq.stack}

The sequential stack sorting problem is in
\cite{Kn73} exercise 5.2.4-20.
We have $k$ stacks numbered $S_0, \ldots, S_{k-1}$. The input
is a permutation $\pi$ of the elements $1, \ldots, n$. 
Initially we push the elements of $\pi$ on $S_0$ at most one at a time
in the order in which they appear in $\pi$. At every step we can pop
a stack (the popped elements will move left in Figure~\ref{stacks}) 
or push an incoming element on a stack. 
The question is how many stack are needed for sorting $\pi$.
It is known that $k = \log n$ stacks suffice, and $\frac{1}{2} \log n$
stacks are necessary in the worst-case~\cite{Kn73,T72}.
Here we prove that the same lower bound also holds
on the average with a very simple incompressibility argument.

\begin{theorem}
On the average, at least $\frac{1}{2} \log n$ stacks are needed for
sequential stack sort. 
\end{theorem}
\begin{proof}
Fix a random permutation $\pi$ such that 
\[
C( \pi | n,P) \leq \log n! - \log = n\log n - O(\log n),
\]
where $P$ is an encoding program to be specified in the following.

Assume that $k$ stacks is sufficient to sort $\pi$. 
We now encode such a sorting process.
For every stack, exactly $n$ elements pass through it. Hence we need perform 
precisely $n$ pushes and $n$ pops on every stack. Encode a push as $0$ and
a pop as $1$. It is easy to prove that different permutations must have
different push/pop sequences on at least one stack.
Thus with $2kn$ bits, we can completely specify the input permutation $\pi$.
\footnote{In fact since each stack corresponds to precisely $n$ pushes
and $n$ pops where the pushes and pops form a ``balanced'' string,
the Kolmogorov complexity of such a sequence is at most 
$g(n) := 2n - \frac{3}{2} \log n + O(1)$ bits. So $2kg(n)$ bits would suffice
to specifiy the input permutation.
But this does not help to nontrivially improve the bound.} 
Then, as before, 
\[ 2kn \geq  \log n! - \log n = n\log n - O(\log n). \]
Hence, approximately $k \geq \frac{1}{2} \log n$ for the random 
permutation $\pi$. 

Since most
permutations are random, we can calculate the average-case lower bound as:
\[ \frac{1}{2} \log n \cdot \frac{n-1}{n} + 1 \cdot \frac{1}{n} \approx
   \frac{1}{2} \log n. \]

\end{proof}

\subsection{Sorting with Parallel Stacks}\label{sect.para.stack}

Clearly, the input sequence $2,3,4, \ldots, n,1$ requires $n-1$
parallel stacks to sort. Hence the worst-case complexity of
sorting with parallel stacks, as shown in Figure~\ref{parallel},
is $n-1$.
However, most sequences do not need this many stacks to sort
in parallel arrangement. The next two theorems
show that on the average, $\Theta (\sqrt{n})$ stacks are both
necessary and sufficient. Observe that the result is actually implied
by the connection between sorting with parallel stacks and 
{\em longest increasing subsequences} given in~\cite{T72} and the bounds
on the length of longest increasing subsequences of random permutations given
in, \cite{king73,LS77,KV77}.
However, the proofs in~\cite{king73,LS77,KV77}
use deep results from probability theory (such as Kingman's ergodic theorem)
and are quite sophisticated. 
Here we give simple proofs using incompressibility arguments.

\begin{theorem}\label{thm.par-stack-upper}
On the average, the number of parallel stacks needed to sort
$n$ elements is $O(\sqrt{n})$.
\end{theorem}
\begin{proof}
Consider a random permutation $\pi$ such that 
\[
C(\pi | n) \geq \log n! - \log n.
\]
We use the following trivial algorithm (which is described in
\cite{T72})
to sort $\pi$ with stacks in the parallel arrangement 
as shown in Figure~\ref{parallel}. 
Assume that the stacks are named $S_0, S_1, \ldots$ and
the input sequence is denoted as $x_1, \ldots , x_n$. 

\bigskip
{\bf Algorithm Parallel-Stack-Sort}
\begin{enumerate}
\item
For $i = 1$ to $n$ do
\begin{description}
\item
Scan the stacks from left to right, and push $x_i$ on the
the first stack $S_j$ whose top element is larger than $x_i$.
If such a stack doesn't exist, put $x_i$ on the first empty stack.
\end{description}
\item
Pop the stacks in the ascending order of their top elements.
\end{enumerate}

We claim that algorithm Parallel-Stack-Sort
uses $O(\sqrt{n})$ stacks on the permutation
$\pi$. First, we observe that if the algorithm uses $m$ stacks on $\pi$ then
we can identify an increasing subsequence of $\pi$ of length $m$ as
in~\cite{T72}. This can be
done by a trivial backtracing starting from the top element of the last stack.
Then we argue that 
$\pi$ cannot have an increasing subsequence of length longer than
$e\sqrt{n}$, where $e$ is the natural constant, since it is
$\log n$-incompressible.

Suppose that $\sigma$ is a longest increasing subsequence of $\pi$ and
$m = |\sigma |$ is the length of $\sigma$. Then we can encode $\pi$ by
specifying:
\begin{enumerate}
\item a description of this encoding scheme in $O(1)$ bits;
\item the number $m$ in $\log m$ bits;
\item the permutation $\sigma$ in $\log {n \choose m}$ bits;
\item the locations of the elements of $\sigma$ in the permutation $\pi$
   in at most $\log {n \choose m}$ bits; and 
\item the remaining $\pi$ with the elements of $\sigma$ deleted in
      $\log (n-m)!$ bits.
\end{enumerate}

This takes a total of
\[ \log (n-m)! + 2\log \frac{n!}{m!(n-m)!} + \log m + O(1) + 2\log\log m \]
bits. Using Stirling approximation and the fact that $\sqrt{n} \leq m = o(n)$,
we can simplify the above expression as:
\begin{eqnarray*}
&& \log (n-m)! + 2\log \frac{n!}{m!(n-m)!} + \log m + O(1) 
   + 2\log\log m \\
& \leq & \log n! + \log \frac{(n/e)^n}{(m/e)^{2m} ((n-m)/e)^{n-m}} + O(\log n) \\
& \approx & \log n! + m\log \frac{n}{m^2} + (n-m) \log \frac{n}{n-m} + m\log e
   + O(\log n) \\
& \approx & \log n! + m\log \frac{n}{m^2} + 2m \log e + O(\log n) 
\end{eqnarray*}

Hence we have inequality
\[ \log n! + m\log \frac{n}{m^2} + 2m \log e + O(\log n) \geq \log n! 
  - \log n \]
which requires that (approximately) $m \leq e\sqrt{n} = O(\sqrt{n})$.

The average complexity of Parallel-Stack-Sort can be simply calculated as:
\[ O(\sqrt{n}) \cdot \frac{n-1}{n} + n \cdot \frac{1}{n} = O(\sqrt{n}). \]
\end{proof}

\begin{theorem}\label{thm.par-stack-lower}
On the average, 
the number of parallel stacks required to sort a permutation 
is $\Omega (\sqrt{n})$.
\end{theorem}
\begin{proof}
Let $A$ be any sorting algorithm using parallel stacks.
Fix a random permutation $\pi$ with 
$C(\pi|n,P) \geq \log n! - \log n$,
where $P$ is the program to do the encoding discussed 
in the following. Suppose that $A$ uses $T$ parallel stacks to sort $\pi$.
This sorting process involves a sequence of moves, and we can
encode this sequence of moves by a sequence of the following terms:
\begin{itemize}
\item
push to stack $i$,
\item
pop stack $j$,
\end{itemize}
where the element to be pushed is the 
next unprocessed element from the input sequence and the popped element
is written as the next output element.
Each of these term requires $\log T$ bits.
In total, we use $2n$ terms precisely
since every element has to be pushed once and popped once.
Such a sequence is unique for every permutation.

Thus we have a description of an input sequence with length 
$2n \log  T $ bits, which must exceed 
$C(\pi|n,P) \geq n \log n - O(\log n)$.
It follows that approximately $T \geq  \sqrt{n} = \Omega(\sqrt{n})$.


We can calculate the average-case complexity of $A$ as:
\[ \Omega(\sqrt{n}) \cdot \frac{n-1}{n} + 1 \cdot \frac{1}{n} = 
   \Omega(\sqrt{n}). \]
\end{proof}

\subsection{Sorting with Parallel Queues}\label{sect.para.queue}

It is easy to see that sorting cannot be done with a sequence
of queues. So we consider the complexity of sorting
with parallel queues. It turns out that all the result in the previous
subsection also hold for queues.

As noticed in \cite{T72}, the worst-case complexity of sorting with
parallel queues is $n$ since the input sequence $n, n-1, \ldots, 1$
requires $n$ queues to sort. 
We show in the next two theorems that on the average,
$\Theta (\sqrt{n})$ queues are both necessary and sufficient.
Again, the result is implied by the connection between sorting with parallel
queues and longest {\em decreasing} subsequences given in~\cite{T72} and
the bounds in~\cite{king73,LS77,KV77} (with sophisticated proofs). 
Our proofs are almost trivial given the proofs in the previous subsection.

\begin{theorem}\label{thm.par-queue-upper}
On the average, the number of parallel queues needed to sort
$n$ elements is upper bounded by $O(\sqrt{n})$.
\end{theorem}
\begin{proof}
The proof is very similar to the proof of Theorem~\ref{thm.par-stack-upper}.
We use a slightly modified greedy algorithm as described in \cite{T72}:

\bigskip
{\bf Algorithm Parallel-Queue-Sort}
\begin{enumerate}
\item
For $i = 1$ to $n$ do
\begin{description}
\item
Scan the queues from left to right, and append $x_i$ on the
the first queue whose rear element is smaller than $x_i$.
If such a queue doesn't exist, put $x_i$ on the first empty queue.
\end{description}
\item
Delete the front elements of the queues in the ascending order.
\end{enumerate}

Again, we can claim that algorithm Parallel-Queue-Sort
uses $O(\sqrt{n})$ queues on any $\log n$-incompressible permutation
$\pi$. We first observe that if the algorithm uses $m$ queues on $\pi$ then
a decreasing subsequence of $\pi$ of length $m$ can be identified, and
we then argue that $\pi$ cannot have a decreasing subsequence of
length longer than $e \sqrt{n}$, in a way analogous to the argument in 
the proof of Theorem~\ref{thm.par-stack-upper}.
\end{proof}

\begin{theorem}\label{thm.par-queue-lower}
On the average,
the number of parallel queues required to sort a permutation
is $\Omega (\sqrt{n})$.
\end{theorem}
\begin{proof}
The proof is the same as the one for Theorem~\ref{thm.par-stack-lower}
except that we should replace ``push'' with ``enqueue'' and ``pop'' with
``dequeue''.
\end{proof}

%
%
%
%
%

\section{Open Questions} 

We have shown that the incompressibility method is a quite useful
tool for analyzing average-case complexity of sorting algorithms.
Simplicity has been our goal. All the proofs and methodology
presented here can be easily grasped and applied, as also demonstrated 
in \cite{BJLV99}, and they can be easily taught in the classrooms to
undergraduate students. 

The average-case performance of Shellsort has been one of the most
fundamental and interesting open problems in the area of algorithm
analysis. The simple average-case analysis we made for 
Insertion Sort (1-pass Shellsort), Bubble
Sort, stack-sort and queue-sort are for the purpose of demonstrating
the generality and simplicity of our 
technique in analyzing many sorting algorithms.
Several questions remain, such as:
\begin{enumerate}
\item
Prove tight average-case lower bound for Shellsort. Our bound is not
tight for $p = 2$ passes. 
\item
For sorting with sequential stacks, 
can we close the gap between $\log n$ upper bound and the
$\frac{1}{2} \log n$ lower bound?
\end{enumerate}

{\small
\section{Acknowledgements}
We thank I. Munro, V. Pratt, and R. Sedgewick for telling us many things
about Shellsort and stack/queue sorts.


}

\begin{thebibliography}{99}

\bibitem{BJLV99}
H. Buhrman, T. Jiang, M. Li, and P. Vit\'anyi,
New applications of the incompressibility method,
submitted to {\em ICALP'99}.

 

\bibitem{IS85}
J. Incerpi and R. Sedgewick,
Improved upper bounds on Shellsort,
{\em Journal of Computer and System Sciences},
31(1985), 210--224.

\bibitem{KV77}
S.V. Kerov and A.M. Versik, Asymptotics of the Plancherel measure
on symmetric group and the limiting form of the Young tableaux,
{\em Soviet Math. Dokl.} 18 (1977), 527-531.

\bibitem{king73}
J.F.C. Kingman, The ergodic theory of subadditive stochastic processes,
{\em Ann. Probab.} 1 (1973), 883-909.

\bibitem{Ko65}
A.N. Kolmogorov,
Three approaches to the quantitative definition of information.
{\em Problems Inform. Transmission}, 1:1(1965), 1--7.

\bibitem{Kn73}
D.E. Knuth, {\em The Art of Computer Programming, Vol.3:
Sorting and Searching}, Addison-Wesley, 1973 (1st Edition),
1998 (2nd Edition).

\bibitem{LiVi93}
M. Li and P.M.B. Vit\'anyi, %
\it An Introduction to Kolmogorov Complexity
and its Applications%
\rm , Springer-Verlag, New York, 2nd Edition, 1997.



\bibitem{LS77}
B.F. Logan and L.A. Shepp, A variational problem for random Young
tableaux, {\em Advances in Math.} 26 (1977), 206-222.

\bibitem{PS65}
A. Papernov and G. Stasevich,
A method for information sorting in computer memories,
{\em Problems Inform. Transmission}, 1:3(1965), 63--75.

\bibitem{PPS92}
C.G. Plaxton, B. Poonen and T. Suel,
Improved lower bounds for Shellsort,
{\em Proc. 33rd IEEE Symp. Foundat. Comput. Sci.}, pp. 226--235, 1992.

\bibitem{Pr72}
V.R. Pratt, {\em Shellsort and Sorting Networks},
Ph.D. Thesis, Stanford University, 1972.

\bibitem{Se96}
R. Sedgewick, Analysis of Shellsort and related algorithms,
presented at the {\em Fourth Annual European Symposium on Algorithms},
Barcelona, September, 1996.

\bibitem{Se97}
R. Sedgewick, Open problems in the analysis of sorting
and searching algorithms, Presented at
{\em Workshop on the Probabilistic Analysis of Algorithms},
Princeton, May, 1997. 

\bibitem{Sh59}
D.L. Shell, A high-speed sorting procedure,
{\em Commun. ACM}, 2:7(1959), 30--32. 

\bibitem{T72}
R.E. Tarjan, Sorting using networks of queues and stacks, 
{\em Journal of the ACM}, 19(1972), 341--346.

\bibitem{Yao80}
A.C.C. Yao, An analysis of $(h,k,1)$-Shellsort,
{\em Journal of Algorithms}, 1(1980), 14--50.

\end{thebibliography}
\end{document}